\pdfoutput=1
\documentclass[11pt]{article}
\usepackage[]{acl}
\usepackage{times}
\usepackage{latexsym}
\usepackage[T1]{fontenc}
\usepackage[utf8]{inputenc}
\usepackage{microtype}
\usepackage{booktabs} 
\usepackage{tabularx} 
\usepackage{hyperref}       
\usepackage{url}            
\usepackage{xurl} 
\usepackage{amsfonts}       
\usepackage{nicefrac}       
\usepackage{microtype}      
\usepackage{xcolor}         
\usepackage{graphicx}       
\usepackage{subfig} 
\usepackage{amsmath}
\usepackage{algorithm}
\usepackage{listings}
\usepackage{pdflscape} 
\usepackage{changepage} 

\title{Dewey Long Context Embedding Model: A Technical Report}

\author{
 \textbf{Dun Zhang\textsuperscript{1}\thanks{$^{\ast}$Dun Zhang is the corresponding author.}}, 
 \textbf{Panxiang Zou\textsuperscript{2}},
 \textbf{Yudong Zhou\textsuperscript{1}}
\\
 \textsuperscript{1}PRIORSHAPE
 \textsuperscript{2}RICHINFO
\\
\small{
 \texttt{infgrad@163.com} \quad
 \texttt{zoupanxiang@richinfo.cn}
}
}

\begin{document}
\maketitle
\begin{abstract}
This technical report presents the training methodology and evaluation results of the open-source dewey\_en\_beta embedding model.
The increasing demand for retrieval-augmented generation (RAG) systems and the expanding context window capabilities of large language models (LLMs) have created critical challenges for conventional embedding models. 
Current approaches often struggle to maintain semantic coherence when processing documents exceeding typical sequence length limitations, significantly impacting retrieval performance in knowledge-intensive applications. 
This paper presents dewey\_en\_beta, a novel text embedding model that achieves excellent performance on MTEB (Eng, v2)\cite{enevoldsen2025mmtebmassivemultilingualtext} and LongEmbed benchmark\cite{zhu2024longembed} while supporting 128K token sequences. 
Our technical contribution centers on chunk alignment training, an innovative methodology that enables the simultaneous generation of localized chunk embeddings and global document-level representations through distillation \cite{zhang2025jasperstelladistillationsota}. 
Information regarding the model release can be found at \url{https://huggingface.co/infgrad/dewey_en_beta}.
\end{abstract}

\section{Introduction}
Text embedding models serve as fundamental components in contemporary information retrieval systems and retrieval-enhanced language models. While recent advancements in semantic representation learning have yielded considerable improvements, practical challenges remain in processing extended textual content. The sequence length capacity of embedding models presents an important technical consideration, as enhanced context processing could potentially improve document comprehension and facilitate more adaptable retrieval implementations.

In this technical report, inspired by late chunking\cite{günther2024latechunkingcontextualchunk} we present a preliminary investigation into chunk-alignment training strategies for text embedding models. This exploratory approach attempts to address the representation learning challenges associated with long-form documents through segment-level feature alignment. Our experimental framework demonstrates the feasibility of processing text sequences up to 128k tokens while preserving performance levels on standard evaluation benchmarks. The current implementation offers two operational modes: a conventional single-vector encoding scheme and an experimental multi-vector variant for extended context processing.

The primary contribution lies in developing a chunk-alignment training method (achieved by knowledge distillation) that attempts to extend existing embedding architectures to longer textual sequences. It should be emphasized that this represents an initial exploration rather than a definitive solution, with the current results suggesting potential pathways for further investigation. Our comparative evaluations indicate relatively balanced performance across conventional benchmarks and preliminary long-context test sets, though we acknowledge significant room for improvement in both theoretical framework and practical implementation.

\section{Training Methodology}

\noindent
\textbf{Selection of Base Model}

After investigating many models, we select ModernBERT-Large\cite{warner2024smarterbetterfasterlonger} as the base model of dewey\_en\_beta.
ModernBERT is a modernized bidirectional encoder-only Transformer model (BERT-style) pre-trained on 2 trillion tokens of English and code data with a native context length of up to 8,192 tokens. ModernBERT leverages recent architectural improvements:
\begin{itemize}
    \item Rotary Positional Embeddings (RoPE)\cite{su2023roformerenhancedtransformerrotary} for long-context support.
    \item Local-Global Alternating Attention\cite{gemmateam2024gemma2improvingopen} for efficiency on long inputs.
    \item Unpadding and Flash Attention \cite{NEURIPS2023_095a6917}\cite{zhang-etal-2024-mgte}\cite{zeng2022boostingdistributedtrainingperformance} for efficient inference.
\end{itemize}
To scale ModernBERT's max length to 128k, we change the \textit{global\_rope\_theta} to 73780400 according to \cite{men2024baseropeboundscontext} and \url{https://spaces.ac.cn/archives/10122}.

\noindent
\textbf{Chunk-Alignment Training}

Our model can generate three types of embeddings:
\begin{enumerate}
    \item CLS embedding: A CLS embedding in a BERT-like model. In our training process, it will learn the teacher model's embedding of the whole text.
    \item Chunk embeddings: The mean embeddings of chunk token embeddings. In our training process, it will learn the teacher model's embedding of each chunk.
    \item Mean embedding: The mean embeddings of all token embeddings (excluding CLS, SEP, and prompt token embeddings). It is a special case of chunk embedding (i.e. the chunk is the whole text).
\end{enumerate}

For a more comprehensive introduction of our model and distillation framework, we make the following definitions:
\begin{itemize}
    \item  $teacher$: a teacher embedding model with the function $encode$ to encode texts to embeddings
    \item  $cls\_embed, chunk\_embed_i$: CLS embedding and chunk embedding of $\text{i}_\text{th}$ chunk in student model(i.e. the model to be trained)
    \item  $cls\_teacher\_embed, chunk\_teacher\_embed_i$ as the whole text teacher embedding and chunk teacher embeddings
    \item  $s_x$: The normalized vector representation of a text $x$ produced by the student model.
    \item  $t_x$: The vector representation of the same text $x$, produced by a teacher model.
    \item  $S_X$: A matrix of normalized vector representations for a batch of text $X$ produced by the student model.
    \item  $T_X$: A corresponding matrix of vector representations for the same batch of text $X$, generated by a teacher model.
\end{itemize}

$cls\_teacher\_embed$ and $chunk\_teacher\_embed_i$ can be obtained by the following equations:
$$ cls\_teacher\_embed = teacher.encode(whole text) $$
$$ chunk\_embed_i = teacher.encode(chunk_i) $$

After getting cls and chunks embeddings of student and teacher model, we calculate $cosine\_loss$ \ref{loss1} and $similarity\_loss$ \ref{loss2} as the final training loss.
\begin{equation}
\label{loss1}
\mathcal{L}_{cosine\_loss}=  \sum_x 1 - s_x \cdot t_x.
\end{equation}

\begin{equation}
\label{loss2}
\mathcal{L}_{similarity\_loss} = MSE(S_{X}S_{X}^T, T_{X}T_{X}^T))
\end{equation}

\begin{figure*}[th]
    \centering
    \begin{center}
        \includegraphics[width=0.75\textwidth]{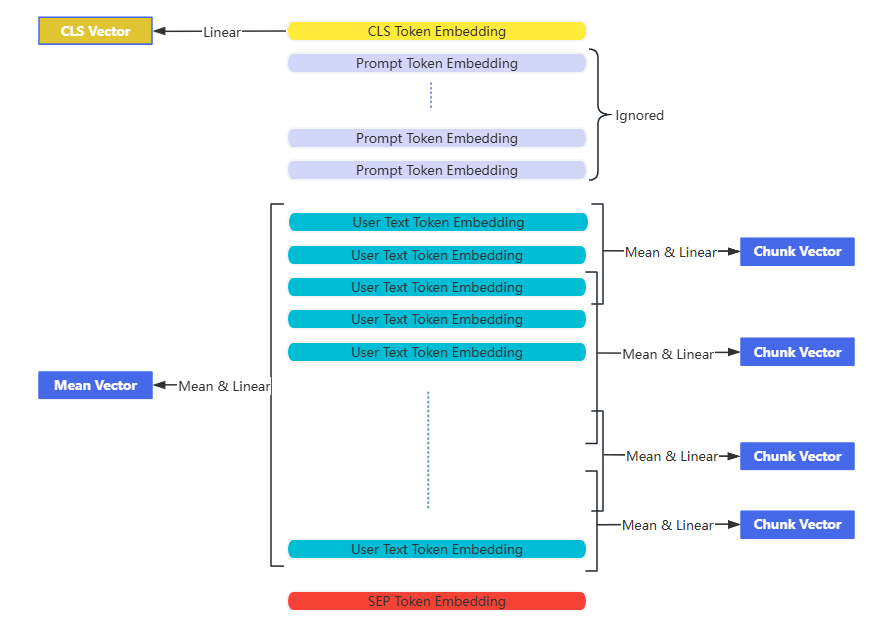}
        \caption{Model architecture}
        \label{fig:model_architecture}
    \end{center}
\end{figure*}

\noindent
\textbf{Implementation Details}

We use Linq-Embed-Mistral\cite{LinqAIResearch2024} as our teacher model. We get unsupervised texts from Infinity-Instruct\cite{InfinityInstruct2024}\cite{zhao2024iidoptimizinginstructionlearning}\cite{zhang2024inifinitymath} and fineweb-edu\cite{lozhkov2024fineweb-edu}. We take two strategies to split text to chunks:
\begin{enumerate}
    \item Split Text by Word
    \item RecursiveCharacterTextSplitter in langchain \url{https://python.langchain.com/docs/introduction/}
\end{enumerate}

We chose to use the RecursiveCharacterTextSplitter with 70\% probability and the Split Text by Word with 30\% probability.
All the two strategies use a randomized chunk\_size(from 64 to 500) and chunk\_overlap(from $0.3*chunk\_size$ to $0.6*chunk\_size$).

We are training our model with about 10 million data (about 100 million chunks). The batch size is set to 64, and the learning rate is 1e-4 with a 2000-step warmup and linear decay.
We use the StableAdamW optimizer\cite{wortsman2023stablelowprecisiontraininglargescale}, which improves upon AdamW\cite{loshchilov2019decoupledweightdecayregularization} by adding Adafactor-style update clipping as a per-parameter learning rate adjustment.
We total train 2 epochs. We set weight decay to zero.
The max length of training data is 2048.

\section{Experimental Results}

\noindent
\textbf{English Text Embedding Benchmark }

MTEB(eng, v2)\cite{enevoldsen2025mmtebmassivemultilingualtext} is a new English Massive Text Embedding Benchmark. This benchmark was created to account for the fact that many models have now been finetuned to tasks in the original MTEB, and contains tasks that are not as frequently used for model training. This way the new benchmark and leaderboard can give our users a more realistic expectation of models' generalization performance.

We evaluated our model's performance on this benchmark. As shown in \ref{tab:mtebresult}, while our model supports a context length of 128k tokens, it still achieves competitive results, outperforming most models of comparable size and even some larger-scale models on this particular benchmark.

\begin{table*}[!ht]
    \centering
    \scalebox{0.5}{
    \begin{tabular}{|l|l|l|l|l|l|l|l|l|l|l|l|l|l|}
    \hline
        Model & Zero-shot & Parameters & Dimensions & Max Tokens & Mean (Task) & Mean (TaskType) & Classification & Clustering & Pair Classification & Reranking & Retrieval & STS & Summarization \\ \hline
        gemini-embedding-exp-03-07 & 95\% & Unknown & 3072 & 8192 & 73.3 & 67.67 & 90.05 & 59.39 & 87.7 & 48.59 & 64.35 & 85.29 & 38.28 \\ \hline
        jasper\_en\_vision\_language\_v1 & 56\% & 1B & 8960 & 131072 & 71.41 & 66.65 & 90.27 & 60.52 & 88.14 & 50 & 56.05 & 84.37 & 37.19 \\ \hline
        gte-Qwen2-7B-instruct & NA & 7B & 3584 & 32768 & 70.72 & 65.77 & 88.52 & 58.97 & 85.9 & 50.47 & 58.09 & 82.69 & 35.74 \\ \hline
        stella\_en\_1.5B\_v5 & 56\% & 1B & 8960 & 131072 & 69.43 & 65.32 & 89.38 & 57.06 & 88.02 & 50.19 & 52.42 & 83.27 & 36.91 \\ \hline
        SFR-Embedding-2\_R & 85\% & 7B & 4096 & 32768 & 69.82 & 65.31 & 90.54 & 59.39 & 88.09 & 48.99 & 53.75 & 80.86 & 35.54 \\ \hline
        Linq-Embed-Mistral & 95\% & 7B & 4096 & 32768 & 69.8 & 65.29 & 83 & 54.07 & 88.44 & 49.44 & 60.14 & 84.69 & 37.26 \\ \hline
        dewey\_en\_beta & 95\% & 395M & 2048 & 131072 & 0.68  & 63.30  & 81.83  & 51.75  & 86.82  & 46.35  & 56.32  & 84.21  & 35.79  \\ \hline
        gte-Qwen2-1.5B-instruct & NA & 1B & 8960 & 32768 & 67.2 & 63.26 & 85.84 & 53.54 & 87.52 & 49.25 & 50.25 & 82.51 & 33.94 \\ \hline
        GritLM-7B & 95\% & 7B & 4096 & 4096 & 67.07 & 63.22 & 81.25 & 50.82 & 87.29 & 49.59 & 54.95 & 83.03 & 35.65 \\ \hline
        GritLM-8x7B & 95\% & 57B & 4096 & 4096 & 66.16 & 62.42 & 79.98 & 51.48 & 85.23 & 49.22 & 52.46 & 82.93 & 35.65 \\ \hline
    \end{tabular}}
    \caption{MTEB(eng, v2) results, rows are sorted in descending order by column Mean (TaskType)}
    \label{tab:mtebresult}
\end{table*}

\noindent
\textbf{LongEmbed Benchmark }

LongEmbed\cite{zhu2024longembed} is a benchmark oriented at exploring models' performance on long-context retrieval. The benchmark comprises two synthetic tasks and four carefully chosen real-world tasks, featuring documents of varying length and dispersed target information.

As can be observed from \ref{tab:longembedresult}, our model achieves reasonably good results with single-vector representation, while the performance is further improved to an optimal level when multi-vector representation is employed.

\begin{table*}[!ht]
    \centering
    \scalebox{0.7}{
    \begin{tabular}{|l|l|l|l|l|l|l|l|}
    \hline
        Model & Zero-shot & Number of Parameters & Embedding Dimensions & Max Tokens & Mean (Task) & Mean (TaskType) & Retrieval \\ \hline
        dewey\_en\_beta-MultiVectors & 100\% & 395M & 2048 & 131072 & 86.59 & 86.59 & 86.59 \\ \hline
        voyage-multilingual-2 & 100\% & Unknown & 1024 & 32000 & 79.17 & 79.17 & 79.17 \\ \hline
        voyage-law-2 & 100\% & Unknown & 1024 & 16000 & 78.85 & 78.85 & 78.85 \\ \hline
        dewey\_en\_beta-SingleVector & 100\% & 395M & 2048 & 131072 & 77.98 & 77.98 & 77.98 \\ \hline
        voyage-3 & 100\% & Unknown & 1024 & 32000 & 74.06 & 74.06 & 74.06 \\ \hline
        inf-retriever-v1 & 100\% & 7B & 3584 & 32768 & 73.19 & 73.19 & 73.19 \\ \hline
    \end{tabular}}
    \caption{LongEmbed results, rows are sorted in descending order by column Mean (TaskType)}
    \label{tab:longembedresult}
\end{table*}

\noindent
\textbf{LoCoV1 Benchmark }

LoCoV1 \cite{saadfalcon2024benchmarkingbuildinglongcontextretrieval}: a novel 12-tasks benchmark constructed to measure long-context retrieval where chunking is not possible or not effective. 

\begin{table*}[!ht]
    \centering
    \scalebox{0.4}{
    \begin{tabular}{|l|l|l|l|l|l|l|l|l|l|l|l|}
    \hline
        dataset-name & bge-m3-8k & gte-modernbert-base-8k & Linq-Embed-Mistral-4k & Linq-Embed-Mistral-8k & SFR-Embedding-Mistral-8k & e5-mistral-7b-instruct-8k & dewey\_en\_beta-8k & dewey\_en\_beta\_64k & dewey\_en\_beta\_64k-multi-vectors \\ \hline
        2wikimqa\_test & 0.9271  & 0.8658  & 0.8884  & 0.9067  & 0.8965  & 0.8901  & 0.8953  & 0.9051  & 0.9775  \\ \hline
        courtlistener\_HTML\_test & 0.1933  & 0.2349  & 0.3551  & 0.3670  & 0.3647  & 0.3543  & 0.3415  & 0.3616  & 0.4775  \\ \hline
        courtlistener\_Plain\_Text\_test & 0.1888  & 0.2478  & 0.3675  & 0.3761  & 0.3679  & 0.3579  & 0.3377  & 0.3485  & 0.4426  \\ \hline
        gov\_report\_test & 0.9869  & 0.9750  & 0.9832  & 0.9837  & 0.9816  & 0.9823  & 0.9855  & 0.9883  & 0.9853  \\ \hline
        legal\_case\_reports\_test & 0.3702  & 0.4476  & 0.5398  & 0.5432  & 0.5319  & 0.4850  & 0.5474  & 0.5875  & 0.6534  \\ \hline
        multifieldqa\_test & 0.9373  & 0.9341  & 0.9345  & 0.9327  & 0.9450  & 0.9321  & 0.9687  & 0.9564  & 0.9754  \\ \hline
        passage\_retrieval\_test & 0.4493  & 0.5271  & 0.3470  & 0.3407  & 0.2902  & 0.3248  & 0.7562  & 0.7389  & 0.8550  \\ \hline
        qasper\_abstract\_test & 1.0000  & 0.9806  & 0.9982  & 0.9982  & 0.9973  & 0.9965  & 0.9973  & 0.9982  & 0.9982  \\ \hline
        qasper\_title\_test & 0.9860  & 0.8892  & 0.9838  & 0.9833  & 0.9861  & 0.9812  & 0.9742  & 0.9742  & 0.9840  \\ \hline
        qmsum\_test & 0.6668  & 0.6307  & 0.6816  & 0.7237  & 0.7169  & 0.7148  & 0.7438  & 0.7613  & 0.8154  \\ \hline
        stackoverflow\_test & 0.9634  & 0.9087  & 0.9760  & 0.9760  & 0.9766  & 0.9690  & 0.9362  & 0.9369  & 0.9443  \\ \hline
        summ\_screen\_fd\_test & 0.9320  & 0.9379  & 0.9747  & 0.9635  & 0.9656  & 0.9580  & 0.9796  & 0.9821  & 0.9788  \\ \hline
        Average & 0.7168  & 0.7150  & 0.7525  & 0.7579  & 0.7517  & 0.7455  & 0.7886  & 0.7949  & 0.8406 \\ \hline
    \end{tabular}}
    \caption{LoCoV1 ndcg@10 results}
    \label{tab:locov1result}
\end{table*}

\section{Conclusion}
In this concise technical report, we present preliminary insights into the dewey model and its training methodology. The proposed approach employs chunk-alignment techniques combined with knowledge distillation, trained on extensive unsupervised data. Our experimental results demonstrate promising performance across both long-text and short-text evaluation benchmarks, though we acknowledge these findings represent early-stage research outcomes.

While observing moderate improvements through chunk-alignment implementation, we recognize substantial room for exploration and refinement in this methodology. This report records current training details and empirical observations, shared with the intention of inviting constructive feedback and collaborative investigation. We hope these preliminary findings might serve as a discussion catalyst within the research community, particularly regarding potential optimizations in alignment strategies and scalability enhancements.

\newpage
\newpage
\bibliography{custom}
\bibliographystyle{acl_natbib}

\end{document}